# ParsRec: A Novel Meta-Learning Approach to Recommending Bibliographic Reference Parsers

Dominika Tkaczyk[1], Rohit Gupta[2], Riccardo Cinti[2] and Joeran Beel[1]

[1] ADAPT Centre, School of Computer Science and Statistics, Trinity College Dublin, Ireland
[2] Iconic Translation Machines Ltd., Dublin, Ireland
d.tkaczyk@gmail.com, rohit@iconictranslation.com,
riccardo@iconictranslation.com, beelj@tcd.ie

**Abstract.** Bibliographic reference parsers extract machine-readable metadata such as *author names*, *title*, *journal*, and *year* from bibliographic reference strings. To extract the metadata, the parsers apply heuristics or machine learning. However, no reference parser, and no algorithm, consistently gives the best results in every scenario. For instance, one tool may be best in extracting titles in ACM citation style, but only third best when APA is used. Another tool may be best in extracting English author names, while another one is best for noisy data (i.e. inconsistent citation styles). In this paper, which is an extended version of [1], we address the problem of reference parsing from a recommender-systems and meta-learning perspective. We propose ParsRec, a meta-learning based recommender-system that recommends the potentially most effective parser for a given reference string. ParsRec recommends one out of 10 open-source parsers: Anystyle-Parser, Biblio, CERMINE, Citation, Citation-Parser, GROBID, ParsCit, PDFSSA4MET, Reference Tagger, and Science Parse. We evaluate ParsRec on 105k references from chemistry. We propose two approaches to meta-learning recommendations. The first approach learns the best parser for an entire reference string. The second approach learns the best parser for each metadata type in a reference string. The second approach achieved a 2.6% increase in F1 (0.909 vs. 0.886) over the best single parser (GROBID), reducing the false positive rate by 20.2% (0.075 vs. 0.094), and the false negative rate by 18.9% (0.107 vs. 0.132).

## 1    Introduction

Bibliographic reference parsing is a well-known task in scientific information extraction and document engineering. In reference parsing, the input is a single reference string, formatted in a specific bibliography style (**Fig. 1**). The output is a machine-readable representation of the input string, typically called a parsed reference (**Fig. 2**). A parsed reference is a collection of metadata fields, each of which is composed of a metadata type (e.g. "year" or "conference") and value (e.g. "2018" or "AICS").

Bibliographic reference parsing is useful for identifying cited documents, also known as citation matching [2]. Citation matching is required for assessing the impact of researchers [3], journals [4, 5] and research institutions [6], and for calculating document similarity [7, 8], in the context of academic search engines [9, 10] and recommender systems [11, 12].

**Fig. 1.** An example bibliographic reference string that could be the input of reference parsing. The marked metadata fields are of types: author name (2 fields), title, journal, volume, issue, year, pages.

```
author:  Acilar, A.M.
author:  Arslan, A.
title:   A collaborative filtering method based on artificial immune network
journal: Expert Systems with Applications
volume:  36
issue:   4
year:    2008
pages:   8324-8332
```

**Fig. 2.** An example of a parsed reference, i.e. a machine-readable representation of the reference string from Fig. 1.

There exist many ready-to-use open-source reference parsers. Recently we compared the performance of ten open source parsers [13]: Anystyle-Parser, Biblio, CERMINE, Citation, Citation-Parser, GROBID, ParsCit, PDFSSA4MET, Reference Tagger and Science Parse. The overall parsing results varied greatly, with F1 ranging from 0.27 for Citation-Parser to 0.89 for GROBID. Our results also showed that different tools have different strengths and weaknesses. For example, ParsCit is ranked 3rd in the overall ranking but is best for extracting author names. Science Parse, ranked 4th overall, is best in extracting the year. These results suggest that there is no single best parser. Instead, different parsers might give the best results for different metadata types and different reference strings. Consequently, we hypothesize that if we were able to accurately choose the best parser for a given scenario, the overall quality of the results should increase. This can be seen as a typical recommendation problem: a user (e.g. a software developer or a researcher) needs the item (reference parser) that satisfies the user's needs best (high quality of metadata fields extracted from reference strings).

In this paper we propose ParsRec, a novel meta-learning recommender system for bibliographic reference parsers. ParsRec takes as input a reference string, identifies the potentially best reference parser(s), applies the chosen parser(s), and outputs the metadata fields. ParsRec is built upon ten open-source parsers mentioned before. ParsRec uses supervised machine learning to recommend the best parser(s) for the input reference string. The novel aspects of ParsRec are: 1) considering reference parsing as a recommendation problem, 2) using a meta learning-based hybrid approach for reference parsing.

This paper is an extended version of a poster published at the 12th ACM Conference on Recommender Systems 2018 (RecSys) [1].

## 2 Related Work

Reference parsers often use regular expressions, hand-crafted rules, and template matching (Biblio [14], Citation [15], Citation-Parser [16], PDFSSA4MET [17], and BibPro [18]). Typically the most effective approach for reference parsing is supervised machine learning, such as Conditional Random Fields (ParsCit [19], GROBID [20], CERMINE [21], Anystyle-Parser [22], Reference Tagger [23] and Science Parse [24]), or Recurrent Neural Networks combined with Conditional Random Fields (Neural ParsCit [25]). To the best of our knowledge, all open-source reference parsers are based on a single technique, none of them uses any ensemble, hybrid or meta-learning techniques.

Some reference parsers are parts of larger systems for information extraction from scientific papers. These systems automatically extract machine-readable information, such as metadata, bibliography, logical structure, or fulltext, from unstructured documents. Examples include PDFX [26], ParsCit [27], GROBID [20], CERMINE [21, 28], Icecite [29, 30] and Team-Beam [31].

Meta-learning is a technique often applied to the problem of algorithm selection [32]. Meta-learning for algorithm selection allows the training of a model able to automatically select the best algorithm for a given scenario. Meta-learning for algorithm selection has been successfully applied to several areas in natural language processing, for example, to grammatical error correction [33], sentiment classification [34], and part-of-speech tagging [35]. To the best of our knowledge, meta-learning has not been applied to reference parsing.

A very effective family of recommender approaches are hybrid-based approaches, which leverage the strengths of many different recommendation algorithms [36]. A weighted hybrid combines the output of many recommenders into one final result [37]. A switching hybrid chooses a single recommender best suited for a given situation [38]. ParsRec can be seen as a switching hybrid of reference parsers, where the switching is controlled by machine learning.

## 3 ParsRec Approach

A meta-learning recommender for reference parsers recommends the best parser for a given scenario. There are multiple ways to define a scenario. One aspect to consider is the granularity of the entity, for which we choose a parser. We can recommend the best parser for:

- a corpus,
- a document, i.e. its bibliography consisting of a list of reference strings,
- a single reference string,
- a metadata type in a reference string, such as title, journal name, or year.

These four parsing levels can also be combined. For example, a recommender system might recommend a parser for a combination of corpus and metadata type. In this case, one parser would be used to extract the year from all reference strings in corpus A, and another parser would be used to extract the names of the authors from all reference strings in corpus B.

In this paper, we examine two types of a meta-learning recommender being inspired by [39]: ParsRec$_{Ref}$ recommends a single parser to an entire reference string, and ParsRec$_{Field}$ recommends a single parser to a pair of reference string and metadata type. The dataset we used for experiments does not allow for other types of the recommender.

ParsRec$_{Ref}$ chooses one parser for a given reference string. This chosen parser is then responsible for the extraction of all metadata. ParsRec$_{Ref}$ works in a few steps (**Fig. 3**). First, for each of the ten parsers, ParsRec$_{Ref}$ predicts the performance of the parser on the given reference string. Second, ParsRec$_{Ref}$ ranks the parsers by their predicted performance. Finally, ParsRec$_{Ref}$ chooses the parser that was ranked highest and applies it to the input reference string.

In ParsRec$_{Ref}$ the prediction of the performance of a parser is done by a linear regression model. We train a separate regression model for every parser. Such a model takes as input the vector of features extracted from the reference string and predicts the F1 that the parser will achieve on this reference string. **Table 1** visualizes the supervised regression problem in ParsRec$_{Ref}$.

For the sake of the machine learning models, the reference strings have to be represented by vectors of features. The features were engineered to capture the citation style and other information that potentially affects the extraction results. We use two types of features: basic heuristics and n-grams.

The heuristics-based features include:

- reference length (1 feature),
- number and fraction of commas (2 features),
- number and fraction of dots (2 features),
- number and fraction of semicolons (2 features),
- whether the reference starts with square bracket enumeration (e.g. "[2]") (1 feature),
- whether the reference starts with dot enumeration (e.g. "14.") (1 feature).

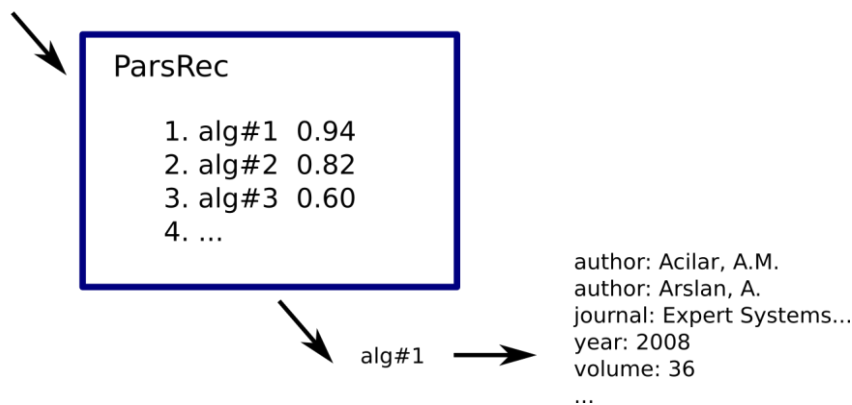

[2] A.M. Acilar, A. Arslan, A collaborative filtering method based on artificial immune network, Expert Systems with Applications 36 (4) (2008) 8324–8332.

ParsRec

1. alg#1  0.94
2. alg#2  0.82
3. alg#3  0.60
4. ...

alg#1 →

author: Acilar, A.M.
author: Arslan, A.
journal: Expert Systems...
year: 2008
volume: 36
...

**Fig. 3.** The workflow of ParsRec_Ref. First, the parsers are ranked based on predicted performance on the input reference string. Second, the parser ranked most highly is chosen and applied to the reference string.

**Table 1.** The visualization of the regression problem in ParsRec_Ref. Each row represents a single reference string. The response variable is the expected F1 of extracted metadata fields. Each parser uses a separate table with the same features and parser-specific response

| | Features | | | | Response |
|---|---|---|---|---|---|
| ref id | ref string length | #commas | bracket enum | … | F1 |
| 1 | 55 | 3 | 0 | … | 0.78 |
| 2 | 78 | 10 | 1 | … | 0.56 |
| 3 | … | … | … | … | … |

N-gram features are binary features corresponding to 3- and 4-grams extracted from the reference string. The terms in n-grams are classes of words, such as *number*, *capitalized word*, *comma*, etc. These features capture style-characteristic sequences of token classes. Example features include: *number-comma-number* (matching e.g. "3, 12"), capitalized *word-comma-uppercase letter-dot* (matching e.g. "Springsteen, B."), *number-left parenthesis-number-right parenthesis* (matching e.g. "5 (28)"). In practice, thousands of distinct n-gram features are generated from the training set, and it is important to select the ones most helpful for the prediction. In our system, we select automatically 150 n-gram features using feature importance, calculated as part of a random forest algorithm trained on the training set [40].

The response variable in the regression model in ParsRec_Ref is the F1 metric. F1 measures how well the metadata fields were extracted from the reference string. F1 is the harmonic mean of precision and recall, calculated by comparing the set of extracted metadata fields to the set of ground truth metadata fields. An extracted field is correct if both type and value are equal to one of the ground truth fields. Precision is the number of correct fields divided by the total number of extracted fields. Recall is the number of correct fields divided by the total number of ground truth fields.

ParsRec$_{Field}$ chooses the potentially best single parser separately for each metadata type in the input reference string. All chosen parsers are then applied to the input reference string. From each parser, the system takes only those metadata fields, for which this parser was chosen. For example, for a specific reference string, ParsRec$_{Field}$ might choose the following parsers: GROBID for extracting authors, title and journal, Science Parse for extracting the year, and CERMINE for volume, issue and pages. In this case, the final metadata fields will contain title field from GROBID, year from Science Parse, etc.

ParsRec$_{Field}$ works in several steps (**Fig. 4**). First, ParsRec$_{Field}$ iterates over all pairs (parser, metadata type), and for each pair ParsRec$_{Field}$ predicts whether the parser will correctly extract the metadata type from the input reference string. Second, for each metadata type, ParsRec$_{Field}$ ranks the parsers based on the predicted probability of being correct and chooses the parser ranked most highly. All chosen parsers are applied to the input reference string and the fields are chosen according to the previous choice of the parser.

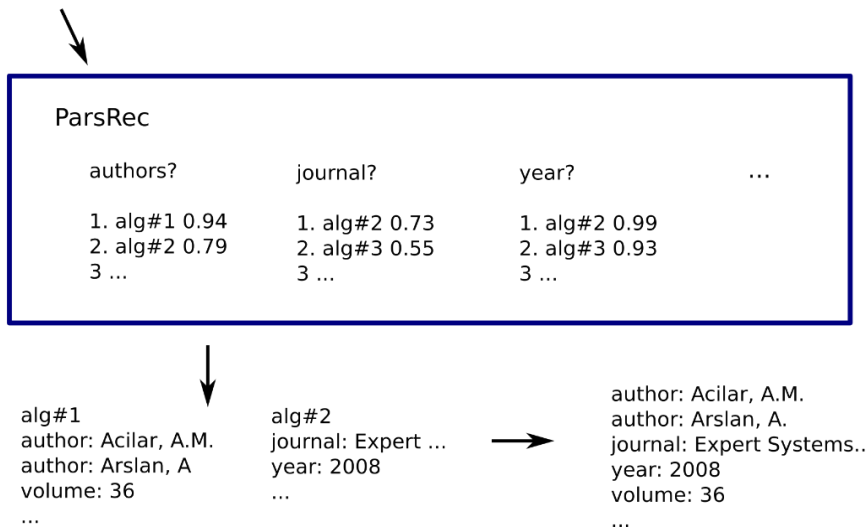

**Fig. 4.** The workflow of ParsRec$_{Field}$. In this case, separate parser rankings are calculated for each metadata field. All parsers ranked most highly are then applied to the input reference string.

In ParsRec$_{Field}$ the prediction of the correctness is done by a binary classifier based on logistic regression. We train a separate classification model for each pair (parser, metadata type). Such a model takes as input the vector of features extracted from the reference string. The features are identical as in the case of ParsRec$_{Ref}$. The model then predicts whether the parser will extract the given metadata field correctly. Apart from a binary classification decision, the logistic regression model outputs the probability of correctness, which is used for ranking. **Table 2** visualizes the classification problem in ParsRec$_{Field}$.

**Table 2.** The classification problem in ParsRec$_{Field}$. Each row represents a single reference string. The response variable corresponds to "is the metadata type extracted correctly". Each pair (parser, metadata type) uses a separate table.

| | Features | | | | Response |
|---|---|---|---|---|---|
| ref id | ref string length | #commas | bracket enum | … | correct? |
| 1 | 55 | 3 | 0 | … | 1 |
| 2 | 78 | 10 | 1 | … | 0 |
| 3 | … | … | … | … | … |

## 4 Methodology

For the experiments we used a closed dataset that comes from a commercial project described in more detail in [13]. The dataset is composed of 371,656 reference strings and the corresponding parsed references, extracted from 9,491 documents from chemical domains. The parsed references were manually curated and contain 1.9 million metadata fields.

The dataset contains 6 metadata types: *author* (the name of the first author), *source* (the source of the referenced document, this can be the name of the journal or the conference, URL or identifier such as arXiv id or DOI), *year*, *volume*, *issue*, and *page* (the first page of the pages range). Unlike the typical reference parsing task, the title of the referenced document was not required by the client of the business project and is not annotated in the data.

The data was randomly split in the following way: 40% of the documents for the training of individual parsers (the training set), 30% of the documents for the training of the parser recommender (the meta-learning set), and 30% of the documents for testing (the test set). Since the split was random, it is possible that there were some rare cases of the same reference string used for both training and testing (if it was contained by two different documents).

The training set will be used in the future for the training of single parsers, to make them work better (this is outside the scope of this paper). The meta-learning set was used for training of the meta-learning recommenders. All parsers were applied to the meta-learning set and evaluated. As a result of the evaluation, we obtained information about which individual metadata fields extracted by the parsers were correct, as well as the overall F1 of each parser on each reference string. This corresponds directly to the data needed for the training of the recommenders (**Table 1** and **Table 2**). Finally, the test set was used for testing and comparisons.

We compare the proposed approach against three baselines. The first baseline is the best single parser (GROBID). The second baseline, called a hybrid baseline, uses the best parser for each metadata type (i.e. ParsCit for author, Science Parse for year, GROBID for other metadata types). The third baseline is a voting ensemble, in which the final result contains only those metadata fields, that appear in the output of at least three different parsers. We evaluate ParsRec in both versions, ParsRec$_{Ref}$ and ParsRec$_{Field}$. We report the results using precision, recall and F1 calculated for the metadata fields.

## 5      Results

The overall results are presented in **Fig. 5**. In general, ParsRec$_{Field}$ achieved the best results, outperforming ParsRec$_{Ref}$ by 2% (F1 0.909 vs. 0.891). This is most likely caused by ParsRec$_{Field}$ being more granular, i.e. it applies parsers separately for different metadata fields, while ParsRec$_{Ref}$ treats reference parsing as a single task.

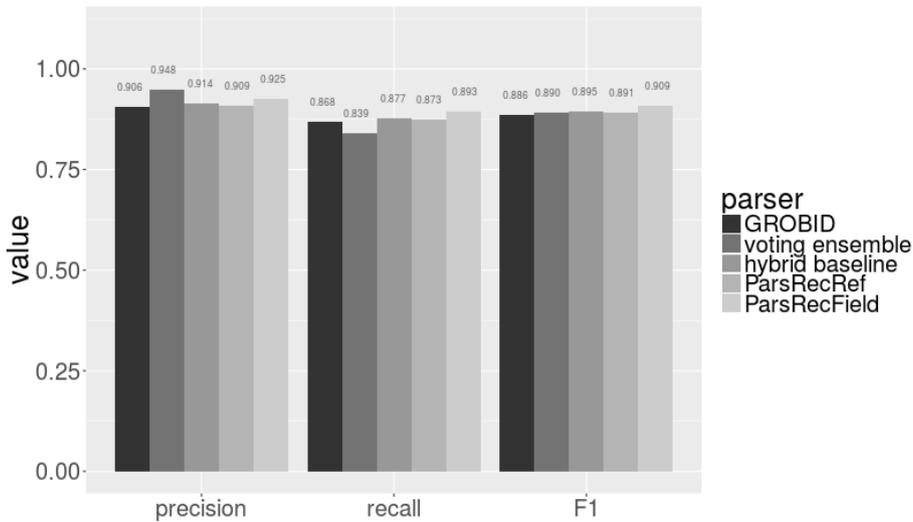

**Fig. 5.** The comparison of the results of three baselines and two variations of ParsRec.

Both variations of ParsRec outperform the best single parser (GROBID). ParsRec$_{Ref}$ achieved a 0.6% increase in F1 (0.891 vs. 0.886), reducing the false positive rate by 3.2% (0.091 vs. 0.095), and the false negative rate by 3.8% (0.127 vs. 0.132). ParsRec$_{Field}$ achieved a 2.6% increase in F1 (0.909 vs. 0.886), reducing the false positive rate by 20.2% (0.075 vs. 0.094), and the false negative rate by 18.9% (0.107 vs. 0.132). We also used Student's t-test to statistically compare the mean F1s over the documents in the test set. Both versions of ParsRec achieved statistically significant increase in mean F1 over GROBID (p = 0.0027 for ParsRec$_{Ref}$ and p < 0.001 for ParsRec$_{Field}$). These improvements show that the recommender indeed learns useful patterns from the data and is able to recommend parsers well.

Both versions of ParsRec also outperform the voting ensemble. While ParsRec$_{Ref}$ is only marginally better (F1 0.890 vs. 0.891), ParsRec$_{Field}$ achieved a 2.1% increase in F1 (0.909 vs. 0.890). In the case of ParsRec$_{Ref}$, the increase in the mean F1 is not statistically significant. In the case of ParsRec$_{Field}$ the increase is significant (p < 0.001).

Only ParsRec$_{Field}$ outperforms the hybrid baseline with a 1.6% increase in F1 (0.909 vs. 0.895). In this case, the increase in the mean F1 is significant (p < 0.001). ParsRec$_{Ref}$ is slightly worse than the hybrid baseline. The reason is most likely the fact that the hybrid baseline is more granular than ParsRec$_{Ref}$.

**Fig. 6** shows how often each parser is chosen in each type of ParsRec. In the case of ParsRec$_{Ref}$, the distribution is more skewed. For example, one the two most often chosen parsers (GROBID and CERMINE) is chosen in 88% of cases in ParsRec$_{Ref}$ and in 65% of cases in ParsRec$_{Field}$. Also, Science Parse, which is almost never chosen in ParsRec$_{Ref}$, is chosen in 8% of cases in ParsRec$_{Field}$. These results show that choosing a parser for different metadata types individually allows for the more effective use of parsers specializing in certain fields, and gives better results.

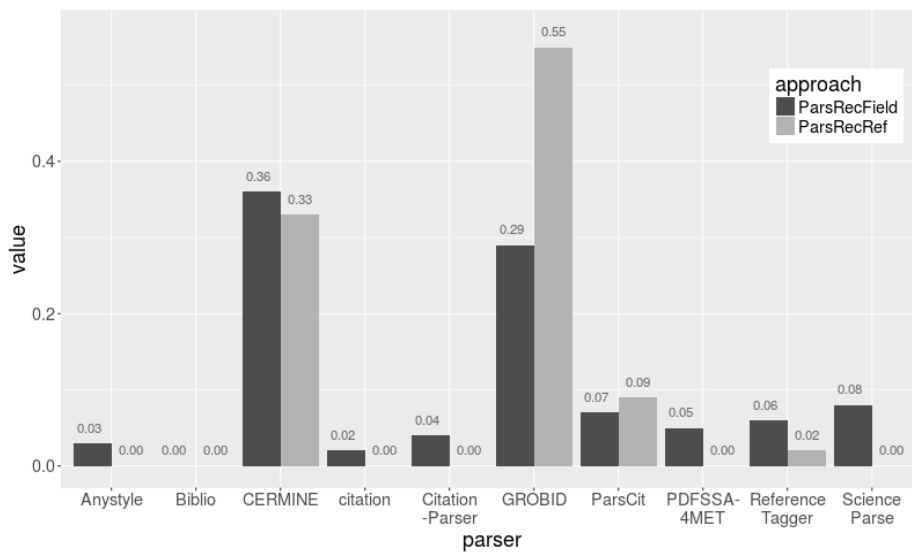

**Fig. 6.** The distributions of recommended parsers in two types of ParsRec.

# 6 Conclusions and Future Work

The promising results of our evaluation clearly show the potential of the proposed recommender system for reference parsers. Both proposed approaches outperform the best single parser and the voting ensemble, which indicates that the recommender indeed makes useful recommendations. One of the proposed approaches (ParsRec$_{Field}$) also outperforms the hybrid baseline.

In most cases, the increases in F1 are not large. We suspect the reason for this is not enough diversity, both in the data and among the parsers. The data comes exclusively from chemical papers, which might not include a lot of different reference styles and languages. Six out of 10 parsers use Conditional Random Fields.

Our plans for the future include training individual parsers, adding more features (related to the language or source of the reference), diversifying the dataset and adding more diverse reference parsers.

# 7 Acknowledgements

This research was conducted in collaboration with and part-funded by Iconic Translation Machines Ltd. with additional financial support from Science Foundation Ireland (SFI) under Grant Number 13/RC/2106. The project has also received funding from the European Union's Horizon 2020 research and innovation programme under the Marie Sklodowska-Curie grant agreement No 713567.